\documentclass[twocolumn,showpacs,preprintnumbers,amsmath,amssymb]{revtex4}

\usepackage{graphicx}
\usepackage{dcolumn}
\usepackage{bm}

\begin{document}


\title{Cold atoms in videotape micro-traps}

\author{C. D. J. Sinclair}
\author{J. A. Retter}
\altaffiliation[Now at ]{Laboratoire Charles Fabry de l'Institut d'Optique, UMR8501 du CNRS, 91403 Orsay Cedex, France.}%
\author{E. A. Curtis}
\email{a.curtis@imperial.ac.uk}
\author{B. V. Hall}
\altaffiliation[Now at ]{Centre for Atom Optics and Ultrafast
Spectroscopy, Swinburne University of Technology, Melbourne,
Australia.}
\author{I. Llorente Garcia}%
\author{S. Eriksson}
\author{B. E. Sauer}%
\author{E. A. Hinds}%

\affiliation{%
Blackett Laboratory, Imperial College, London, SW7 2BW, UK\\
}%

\date{\today}

\begin{abstract}

We describe an array of microscopic atom traps formed by a pattern
of magnetisation on a piece of videotape.  We describe the way in
which cold atoms are loaded into one of these micro-traps and how
the trapped atom cloud is used to explore the properties of the
trap. Evaporative cooling in the micro-trap down to a temperature of
$1\,\mu$K allows us to probe the smoothness of the trapping
potential and reveals some inhomogeneity produced by the magnetic
film.  We discuss future prospects for atom chips based on
microscopic permanent-magnet structures.

\end{abstract}

 \maketitle

\section{Introduction}\label{sec:intro}
Magnetic storage media provide a convenient way to prepare small
magnetic structures that can be used to manipulate cold atoms.
These include audiotape~\cite{Roach}, floppy
disks~\cite{Hughes1,Hughes2,Saba},
videotape~\cite{Rosenbusch1,Rosenbusch2}, magneto-optical
films~\cite{Hannaford,Eriksson,Spreeuw}, and hard
disks~\cite{Lev}. Using standard lithographic techniques, it is
also possible to make small patterns of current carrying wires for
the same purpose~\cite{Folman,ReichelReview}. This has led to the
concept of the atom chip, in which small structures are integrated
on a substrate and used to control the motion and interactions of
microscopic cold atom clouds. The atom chip presents opportunities
for quantum coherent manipulation of atoms for quantum information
processing or fundamental studies of quantum gases. It also
creates a versatile miniaturised system ideal for manipulating
cold atom clouds in a variety of applications such as
magnetometry~\cite{Schwindt},
interferometry~\cite{HindsInterferometer} and miniature
clocks~\cite{Reichelatomclock,Knappe}. For a number of years it
has been clear that permanent-magnet microstructures have a
significant contribution to make to atom chips~\cite{HindsReview},
since they have the possibility of making very tight atom traps
with cloud sizes approaching 10\,nm. So far, however, magnetic
microstructures have only been used to control  much larger
macroscopic clouds.

In this paper we describe a permanent-magnet atom chip, an
adaptation of the magnetic mirror described in
reference~\cite{HindsReview}, in which we prepare and manipulate
small atom clouds with radial sizes down to 50\,nm. The next
section introduces the videotape and shows how periodic
magnetisation can be used to trap atoms. Section
\ref{sec:theatomchip} describes the design and structure of the
videotape atom chip. The loading, cooling and trapping of atoms in
a videotape micro-trap are described in Section \ref{sec:loading}.
In Section \ref{sec:experiments} we describe experiments to test
our understanding of these micro-traps and to study the quality of
the videotape. In the final section we discuss some of the future
directions of permanent-magnet atom-chip experiments.

\section{Magnetic media and atom traps}
\label{sec:magmedia}

\begin{figure}
\resizebox{0.85\columnwidth}{!}{%
 \includegraphics{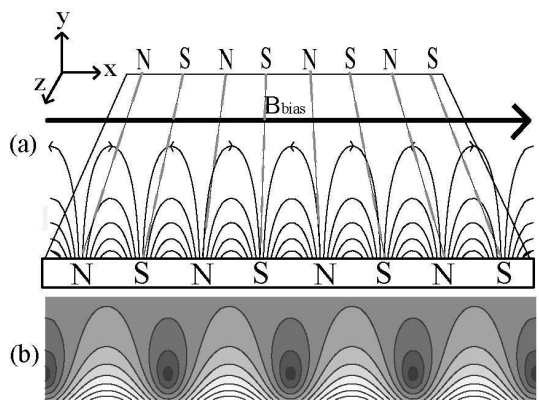}
    }
 \caption{(a) Magnetic field lines produced by the
 magnetised videotape. A uniform bias field is added to this to make
 an array of atom guides. (b) Contours of constant magnetic field
strength. Circular contours enclose the lines of minimum field
strength, where atoms are trapped.}
\label{fig:fieldlinesblack}       
\end{figure}

The basis of our micro-traps is a magnetic pattern written on
videotape as illustrated in Fig.~\ref{fig:fieldlinesblack}. As in
the magnetic mirror we use a sinusoidal pattern, however, in this
work we add a bias field to form an array of micro-traps. With
magnetisation $M_0\cos{k x}$ along the $x$ direction, the magnetic
field outside the film is
\begin{equation}\label{equ:magfield}
(B_x,B_y) = B_{sur} e^{-ky}(-\cos(kx),\sin(kx)).
\end{equation}
For a film of thickness $b$, the field at the surface is
\begin{equation}\label{equ:surfield}
B_{sur} = \frac{1}{2} \mu_{0} M_0(1 - e^{-kb}).
\end{equation}
Thus the potential energy of a weak-field-seeking atom increases
exponentially as it approaches the mirror surface. This is how the
atom is reflected.

The micro-traps appear when a bias field $B_{bias}$ is superimposed
in the $x$-$y$ plane, as shown in Fig.~\ref{fig:fieldlinesblack}.
Near each trap the magnetic field has a quadrupole structure with a
zero in the centre and a gradient of field strength given by
\begin{equation}
B^{\prime} = k B_{bias}.
\end{equation}
In order to suppress Majorana spin flips of the trapped atoms, we
also apply an axial bias field $B_z$, which prevents the total
field from going exactly to zero at the centre. For
small-amplitude transverse oscillations this makes a harmonic trap
with frequency
\begin{equation}\label{equation:omega}
2 \pi f_{r} = k B_{bias}\sqrt{\frac{\mu_{B} g_F m_F}{m B_{z}}},
\end{equation}
where $\mu_{B} g_F m_F$ is the usual factor in the Zeeman energy and
$m$ is the mass of the atom.  The trap centres are formed at a
distance $y_{trap}$ from the surface, given by
\begin{equation}\label{equation:y(B)}
y_{trap} =\frac{1}{k} \ln(B_{sur}/B_{bias}).
\end{equation}
Ref.~\cite{HindsReview} provides further detail about permanent
magnetic patterns and their corresponding fields.

\section{Videotape atom chip design}
\label{sec:theatomchip}

\begin{figure}
\resizebox{0.9\columnwidth}{!}{
   \includegraphics{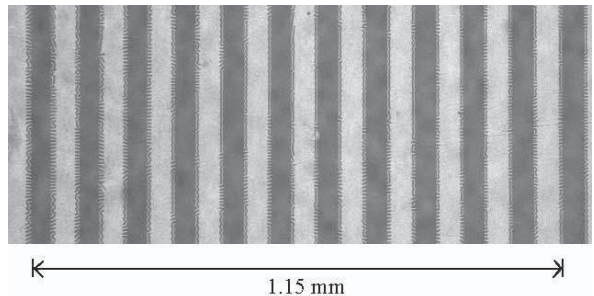}}
 \caption{Optical microscope image showing reversals of the magnetic
 field direction. These are revealed by a thin garnet film placed on
 the surface, viewed with polarised light through a crossed analyser.}
\label{fig:wavelength}
\end{figure}

We have chosen to use videotape because it can store patterns with
feature sizes down to a few micrometers using simple commercial
recording equipment adapted in the laboratory. Videotape is
designed to hold data reliably for long periods of time and has a
high coercivity, making the magnetisation insensitive to the
presence of the bias fields. The particular tape we use, Ampex 398
Betacam SP, has a 3.5$\,\mu$m-thick magnetic layer containing
iron-composite needles, 100 nm long with 10\,nm radius, which are
set in a glue and aligned parallel to the $x$ direction. This film
is supported by a polymer ribbon 11\,$\mu$m thick. Since atom-trap
experiments are conducted in ultra-high vacuum conditions at a few
times $10^{-11}$\,Torr, the vacuum properties of the materials we
use are important. Remarkably, this videotape has a very low
outgassing rate at room temperature and is able to withstand
baking at 120$^\circ$C~\cite{Hopkins}.

\begin{figure}
\resizebox{1\columnwidth}{!}{%
  \includegraphics{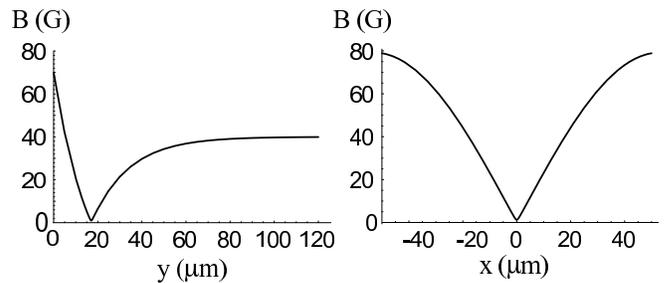}
}
 \caption{Magnetic field strength plotted versus $x$ and $y$ going through
 the centre of a trap. Here we take bias fields of
40\,G along $x$ and 1\,G along $z$. The videotape field has a
strength of 110\,G at the surface and the trap forms 17$\,\mu$m
away from the surface.} \label{fig:VTPotentials}
\end{figure}

For the experiments described in this paper, the magnetic sine wave
has a period of 106\,$\pm$\,2\,$\mu$m. We are able to observe this
directly by placing a thin Faraday-rotating garnet film (Nd:Lu:Bi)
on the surface and viewing it under a microscope with polarised
light, as shown in Fig.~\ref{fig:wavelength}. The accurate
measurement of the period is achieved by counting a large number of
these fringes.

We have measured that the field at the surface of the tape is
$110\pm10$\,G, as we discuss further in
Section~\ref{sec:experiments}. Taking this value together with
typical bias fields $B_{bias}=40$\,G and $B_{z}=1$\,G, we
calculate the expected trapping field strength shown in Fig.
\ref{fig:VTPotentials}. For trapped $^{87}$Rb atoms in the
$|F=2,m_{F}=2\rangle$ sublevel of the ground state, the
small-amplitude radial oscillation frequency is 30\,kHz.

\begin{figure}
\resizebox{0.8\columnwidth}{!}{%
  \includegraphics{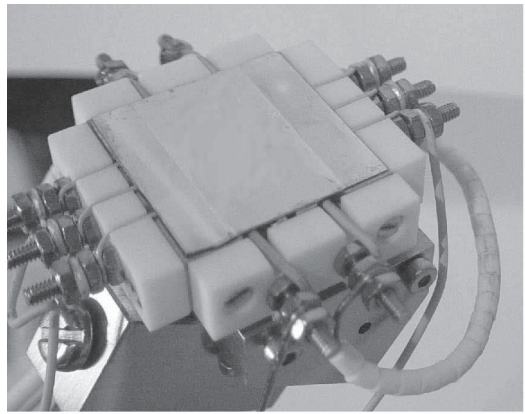}
}
 \caption{The videotape atom chip assembly.  The reflective surface of the chip is one inch square.} \label{fig:atomchip}
\end{figure}

After recording, a piece of the videotape is glued to a glass
coverslip $150\,\mu$m thick using a UHV-compatible epoxy (BYLAPOX
7285), as described in references~\cite{BertramPRA,Retter}. This is
coated with a $\sim$ 5\,nm adhesion layer of chromium and $\sim
400$\,nm of gold in order to make the surface of the chip reflect
780\,nm laser light. The purpose here is to allow the formation of a
magneto-optical trap (MOT) by reflection\,\cite{ReichelHaensch} in
order to collect and cool the atoms close to the videotape. The
coverslip is then glued to a 1\,inch square stainless-steel base to
form the chip assembly shown in Fig.~\ref{fig:atomchip}. Five wires
insulated by a ceramic coating run underneath the coverslip through
channels cut into the steel, ending on Macor terminal blocks. The
three wires parallel to the $z$ direction lie immediately below the
coverslip and have 500\,$\mu$m diameters. The central one of these,
the ``centre wire'', is used to transport cold atoms from the MOT
into one videotape micro-trap, whilst the outer two form an rf
antenna used subsequently for evaporative cooling of the atoms. Two
larger, 1\,mm-diameter wires run along $x$ and pass below the centre
wire. These are separated by 8.5\,mm. Currents in these allow the
trap to be closed off at its ends by making a field that rises to a
maximum in the $z$ direction above each wire. The whole assembly is
mounted in the high-vacuum chamber on a flange that allows
translation under vacuum in any direction.

\section{Loading a micro-trap}
\label{sec:loading}

A double MOT system delivers cold $^{87}$Rb atoms to the atom chip
assembly. The first MOT, in an auxiliary chamber, is a low-velocity
intense source (LVIS)\cite{Lu} with three retro-reflected laser
beams. The Rb atoms are provided by a resistively-heated dispenser
that increases the pressure to $\sim 1\times 10^{-9}$\,Torr. The
LVIS delivers a $\sim 15$\,m/s beam of cold atoms into the main
vacuum chamber through a 6\,mm-long, 1\,mm-diameter aperture 30 cm
away from the atom chip. The second MOT, in the main chamber, is
formed by four independent laser beams, two of which we reflect from
the surface of the chip. The magnetic quadrupole field is provided
by coils located outside the chamber. When this MOT is loaded by the
LVIS source for 20\,s, $\sim3\times 10^8$ atoms are collected at a
temperature of $\sim150\,\mu$K and at a distance of 4\,mm from the
chip. The pressure in this chamber is too low to register on the
ionisation gauge, which cuts out below $3\times10^{-11}$ Torr. The
MOT lifetime exceeds 50\,s.

Once the MOT has been filled, the dispenser and laser beams of the
LVIS are turned off and the cloud is moved to 1.5\,mm from the chip
by ramping up an external bias field of $\sim4$\,G. At this point,
the red-detuning of the laser from the D2($F=2 \rightarrow 3$)
cooling transition is increased over 14\,ms from -15\,MHz to
-45\,MHz. This cools the cloud to $\sim50$\,$\mu$K. Next, the MOT is
switched off (both the light and the field) and a uniform field is
applied in order to optically pump using $\sigma^+$ light on the
D2($F=2 \rightarrow 2$) transition for 400\,$\mu$s. This transfers
the atoms into the $|F=2,m_F=2\rangle$ weak-field-seeking state.
Approximately $3 \times 10^{7}$ of these atoms are then recaptured
in a purely magnetic trap, formed by passing 15\,A through the
centre wire and imposing a 14\,G field along $x$.  The efficiency of
transfer from the MOT to the magnetic trap is particularly sensitive
to the way in which the laser frequency is swept to larger detuning.

The bias field is immediately ramped up to 44 G over 100 ms, which
increases the trap depth, compresses the cloud and moves it to
160\,$\mu$m from the surface (where the field of the videotape is
still negligible). This transfer heats the cloud adiabatically,
increasing the elastic collision rate to $\sim20$\,s$^{-1}$, which
is sufficient to start evaporative cooling. The field at the centre
of this trap is $\sim500$\,mG, giving a radial oscillation frequency
of 1.1\,kHz, while the axial frequency is 15\,Hz. In order to drive
the evaporation, the rf field is now applied and its frequency is
swept for 6 seconds according to $f=(30\,\mbox{e}^{-t}+3.9)\,$MHz,
where $t$ is in seconds. This cools the cloud to 10 $\mu$K. Leaving
the rf field on at 3.9\,MHz, we continue to evaporate by reducing
the centre-wire current and the bias field over 4\,s. This gradually
brings the cloud close to the surface, into a region where the field
of the videotape becomes comparable with the bias field. Finally,
the wire current is ramped down to zero, leaving atoms confined
solely by the videotape micro-trap. After this phase of compression
and additional cooling there are a few times $10^5$ atoms in the
micro-trap at a temperature of $\sim10\,\mu$K. When lower
temperatures than this are required, we make a second stage of rf
evaporation that takes place entirely in the micro-trap.

\begin{figure}
\resizebox{0.9\columnwidth}{!}{%
  \includegraphics{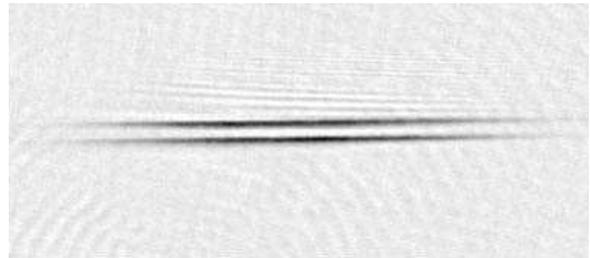}
}
 \caption{Double image of atoms in a single videotape micro-trap. Dark
areas correspond to regions of maximum absorption.  Image area is
3.4\,mm$\times$1.5\,mm.} \label{fig:intrap}
\end{figure}

The atoms are imaged by optical absorption, using a single pulse
of light (typically $40\,\mu$s long), tuned to the
D2($F=2\rightarrow 3$) transition. A CCD camera records the image
of the cloud, formed at unit magnification by a high-quality
achromatic doublet lens. The imaging beam makes an angle of
$\sim10^\circ$ to the gold surface in order to give two images:
the cloud and its reflection in the mirror. This provides a simple
measure of the distance between the atoms and the surface. Figure
\ref{fig:intrap} shows the image of $2\times10^5$ atoms at
$20\,\mu$K, held in a 15\,Hz$\times$4.7\,kHz trap, $55\,\mu$m from
the surface of the videotape.

\section{Experiments}
\label{sec:experiments}

\begin{figure}
\resizebox{0.9\columnwidth}{!}{%
  \includegraphics{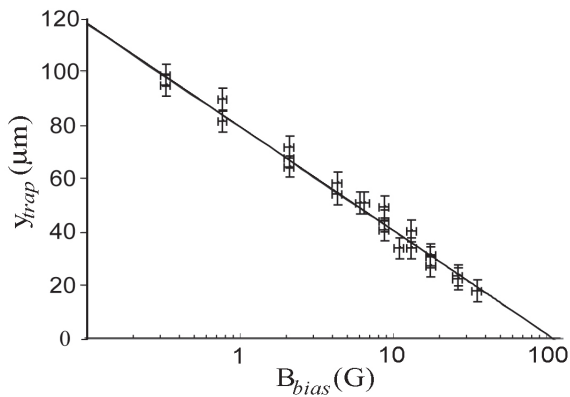}
}
 \caption{Logarithmic dependence of atom-surface distance on the applied bias
field. The intercept at zero height measures the field at the
surface of the chip.} \label{fig:heightvsB}
\end{figure}

Having now trapped the atoms, our first experiment is to check the
exponential law for the strength of the videotape field given in
Eq.(\ref{equ:magfield}) and to measure the surface field $B_{sur}$.
According to Eq.(\ref{equation:y(B)}), the distance between the trap
and the surface, $y_{trap}$, should be linearly proportional to
$\ln(B_{bias})$, with a slope of $1/k$ and an intercept on
$y_{trap}=0$ of $B_{bias}=B_{sur}$. Figure~\ref{fig:heightvsB} shows
measurements of $y_{trap}$ for a variety of bias fields. These
distances, to which we assign an uncertainty of $\pm$\,4\,$\mu$m are
derived from the separation measured between the centres of the two
cloud images. The uncertainty in $B_{bias}$ is a 7\% calibration
error. Our measurements confirm the exponential field decay of
Eq.(\ref{equ:magfield}) and yield a value of $110\pm 10\,$G for
$B_{sur}$. This is approximately half the maximum possible value
given by Eq.(\ref{equ:surfield}) with $\mu_{0} M_0$ equal to the
saturated magnetisation, 2.3\,kG. The result is not surprising since
our method of recording was to monitor the playback signal and to
set the current in the record head just below the level where any
distortion was evident.

\begin{figure}
\resizebox{\columnwidth}{!}{
\includegraphics{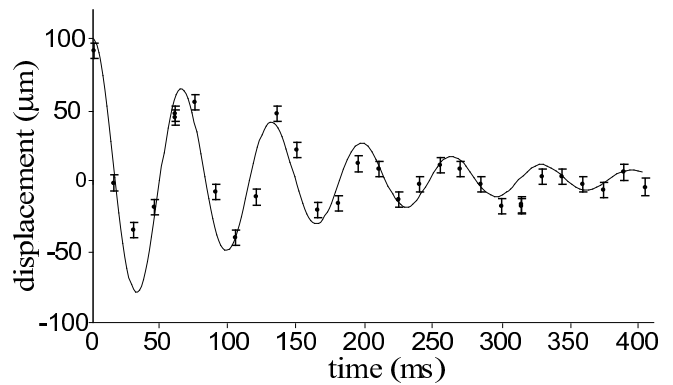}}
\caption{Oscillations of the atom cloud in the axial direction. Each
point involves a new realization of the experiment with a suitably
chosen time delay before measuring the position of the cloud. The
line is a damped sine wave intended to guide the eye.}
\label{fig:axialosc}
\end{figure}

The second experiment is to investigate the frequencies of the
micro-trap. In the axial direction, the restoring force is small
enough that we can see oscillations of the cloud directly on the
CCD camera. Typical axial frequencies are in the range $5-20\,$Hz.
Figure\,\ref{fig:axialosc} shows the centre-of-mass oscillation of
a $1\,\mu$K cloud with an initial displacement of $100\,\mu$m. The
damping of this oscillation is due primarily to the collisional
coupling into other degrees of freedom, with relatively little
effect from the slight anharmonicity of the trap.

\begin{figure}
\resizebox{1\columnwidth}{!}{
     \includegraphics{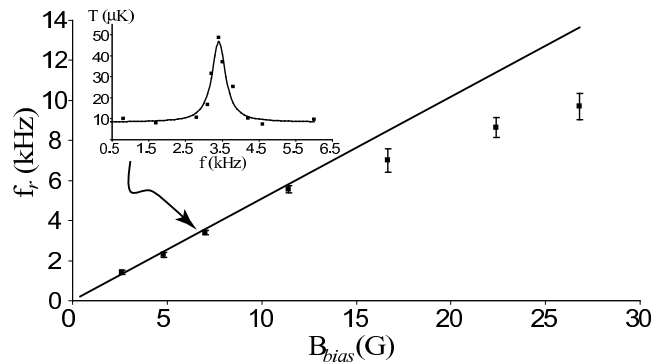}
} \caption{Radial trap frequency versus applied bias field. Solid
line: harmonic approximation given in Eq.~\ref{equation:omega}.
Data points: Measured radial trap frequencies. Inset: Typical
radial excitation spectrum.} \label{fig:radialosc}
\end{figure}

Radial oscillation cannot be observed in the same direct way because
the frequency is too high. Moreover, the radial size of the cloud is
far smaller than one pixel of the camera: a 30\,kHz trap at
$1\,\mu$K, gives a cloud radius of 52\,nm. Instead, we add a
modulation $\delta B_{bias}\cos(2\pi f t)$ to the bias field, which
shakes the trap in the radial direction. When the frequency $f$
coincides with the radial oscillation frequency $f_r$, the cloud is
heated. Once again, the collisions within the gas couple the
motional degrees of freedom, allowing us to detect the radial
heating through the corresponding increase in the length of the
cloud. The inset in Fig.~\ref{fig:radialosc} shows the cloud
temperature versus $f$, as measured by its length after applying the
modulation for $5\,$s. The radial resonance is seen as a dramatic
increase in the temperature. The main graph in
Fig.\,\ref{fig:radialosc} shows these resonance frequencies versus
bias field. The solid line is the harmonic frequency given by
Eq.\,\ref{equation:omega} with $B_z=2\,$G, corresponding to the
axial field used in these experiments. At low bias field, we see
good agreement between this formula and the measured radial
frequency. With a stronger bias, the radial frequency lies below the
prediction of Eq.\,\ref{equation:omega} (and the resonances are
broader). We believe this is due to the compression of the cloud,
which occurs when $B_{bias}$ is increased after loading the
micro-trap. This increases the initial temperature and expands the
cloud out of the central region where the harmonic approximation is
valid. Here the potential is more linear, as one can see in
Fig.\,\ref{fig:VTPotentials} (indeed, the harmonic region is almost
invisible on the scale of Fig.\,\ref{fig:VTPotentials}), giving rise
to a lower oscillation frequency and a broadening of the resonance
feature.

\begin{figure}
\resizebox{1\columnwidth}{!}{
        \includegraphics{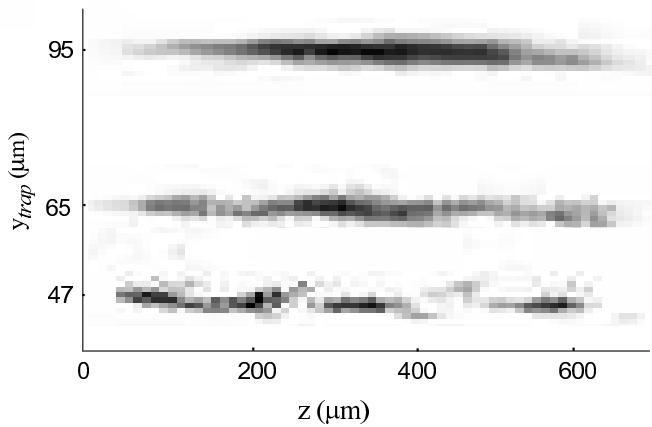}}
 \caption{Density distribution of trapped atom clouds at three
 different heights above the videotape.}
\label{fig:frag}
\end{figure}

Our final experiment uses the cold atoms to probe the magnetic
smoothness of the videotape. Figure \ref{fig:frag} shows the
density distribution of a $1\,\mu$K cloud at three different
distances from the surface. The cloud has a smooth profile at a
distance of $100\,\mu$m, but develops structure as it approaches
the surface. This is reminiscent of the structure observed near a
wire (see~\cite{Fortagh}, \cite{Jones2}, and references therein),
which is caused by an alternating magnetic field in the $z$
direction due to a transverse component of the current density in
the wire. The structure we see here is also caused by an
alternating magnetic field along the $z$ direction, although of
course there are no currents flowing in the videotape. In this
case, it is an alternating component of magnetization along $z$
that causes the unwanted magnetic field through an effective
current density $j_x=\partial M_z/\partial y$.

\begin{figure}
\resizebox{1\columnwidth}{!}{\includegraphics{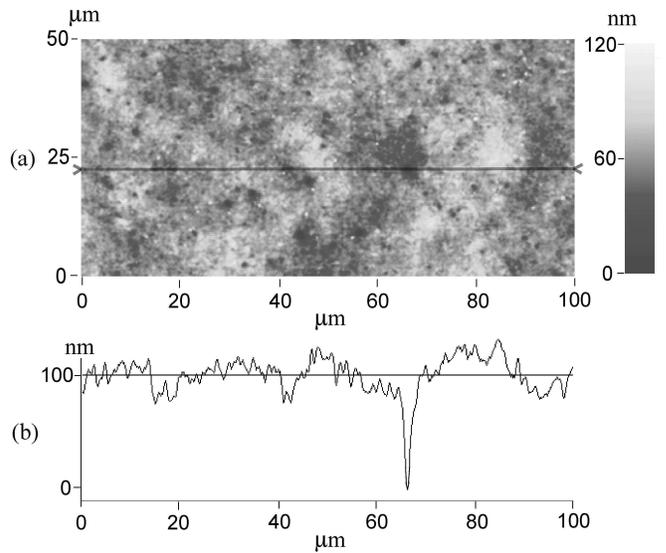}}
\caption{Atomic force microscope scan showing topography of a piece
of videotape. (a) Survey of a $100\,\mu$m$\times\,50\,\mu$m region.
(b) Height versus position along the line indicated in (a).}
\label{fig:AFM}
\end{figure}

In order to discover how this fluctuating $z$ component of
magnetisation comes about, we have studied a piece of tape taken
from the same cassette as the tape. Figure~\ref{fig:AFM} shows the
surface topography of this sample as measured by an atomic force
microscope. The greyscale indicates the height of the tape measured
over an area of $100\,\mu$m$\times\,50\,\mu$m. Below that is a graph
showing the typical height profile along a line. These images reveal
some small height variations over transverse lengths of tens of
micrometers, which are only due to the failure of the tape to lie
flat. In addition, however, we see some small, deep holes in the
surface that only appear on the magnetically coated side of the
tape. The typical spacing of these defects is $\sim10\,\mu$m, with
the more prominent ones being $50-100\,\mu$m apart. It seems most
likely that these are formed during manufacture, perhaps by small
bubbles emerging from the glue. Whatever their cause, such defects
in the magnetic film would be expected to lead to precisely the kind
of structure that we observe in our cold atom clouds.

\section{Conclusion and Prospects}
\label{sec:conclusion}

We have shown how to load cold atoms into micro-traps formed by
patterns of magnetisation on videotape and we have used these clouds
to demonstrate a basic understanding of the magnetic field and the
traps above the tape. The videotape is well suited to making very
anisotropic traps, which are of interest in the study of
one-dimensional quantum gases. Indeed, we have made traps with an
aspect ratio as large as $40\,$kHz$\times\,4\,$Hz. With this in
mind, we have cooled the atom clouds down to $1\,\mu$K and have
studied the smoothness of the trapping potentials formed by the
videotape. This experiment shows no undesirable structure within the
trap at a distance of $100\,\mu$m (where $f_r\leq 1\,$kHz), but
reveals microscopic wells of order $1\,\mu$K deep at a distance of
$50\,\mu$m (where $f_r\leq 10\,$kHz). We have identified that this
is due to physical defects in the magnetic film itself, which
probably render the videotape unsuitable for the study of 1D gas in
smooth traps with the highest radial frequencies. However, the
method of writing magnetic patterns on videotape remains a versatile
and promising technique for manipulating the atoms $100\,\mu$m away
from the surface. At present we are installing optical fibres on the
chip, which will operate at that height and will be used to
investigate the interaction of light with trapped Bose-Einstein
condensates \footnote{A paper reporting Bose-Einstein condensation
on this atom chip is in preparation.}. Further information about the
development of micro-optical components can be found in
\cite{Eriksson2}.

Recent research on Pt/Co multilayer thin films~\cite{Eriksson}
appears to offer a solution to the problem of homogeneity of the
magnetic layer. A pattern of lines with $2\,\mu$m period has been
written on this material and the result indicates that the field
should be exceedingly smooth at heights above a few micrometers.
With these films we anticipate using the techniques developed in
this paper to prepare one-dimensional gases in extremely elongated
traps.

In conclusion, we have shown that cold atoms can be loaded into
micro-traps formed above a videotape, thereby demonstrating a
first rudimentary permanent-magnet atom chip.

\begin{acknowledgments}
We are indebted to Alan Butler, Jon Dyne and Bandu Ratnasekara for
technical assistance.  This work is supported by the UK
Engineering and Physical Sciences Research Council, the Royal
Society, and  the FASTnet and QGates networks of the European
Union.
\end{acknowledgments}

\bibliography{BibEPJ}

\end{document}